# Size-dependent phase transitions in MoS$_2$ nanoparticles controlled by a metal substrate

Albert Bruix, Jeppe V. Lauritsen, Bjørk Hammer*

Interdisciplinary Nanoscience Center (iNANO) and Department of Physics and Astronomy, *Aarhus University, DK-8000 Aarhus C, Denmark*

*E-mail: hammer@phys.au.dk

**Table of Contents Graphic**

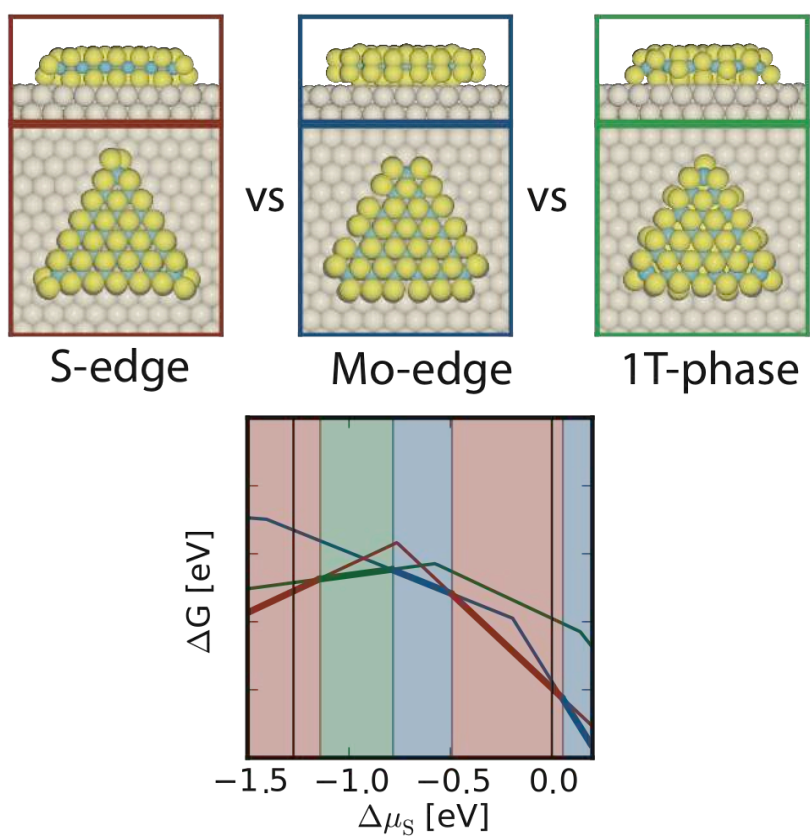

**Abstract**

Nanomaterials based on $MoS_2$ are remarkably versatile; $MoS_2$ nanoparticles are proven catalysts for processes such as hydrodesulphurization and the hydrogen evolution reaction, and transition metal dichalcogenides in general have recently emerged as novel 2D components for nanoscale electronics and optoelectronics. The properties of such materials are intimately related to their structure and dimensionality. For example, only the edges exposed by $MoS_2$ nanoparticles (NPs) are catalytically active, and extended $MoS_2$ systems show different character (direct or indirect gap semiconducting, or metallic) depending on their thickness and crystallographic phase. In this work, we show how particle size and interaction with a metal surface affect the stability and properties of different $MoS_2$ NPs and the resulting phase diagrams. By means of calculations based on the Density Functional Theory (DFT), we address how support interactions affect $MoS_2$ nanoparticles of varying size, composition, and structure. We demonstrate that interaction with Au modifies the relative stability of the different nanoparticle types so that edge terminations and crystallographic phases that are metastable for free-standing nanoparticles and monolayers are expressed in the supported system. These support-effects are strongly size-dependent due to the mismatch between Au and $MoS_2$ lattices, which explains experimentally observed transitions in the structural phases for supported $MoS_2$ NPs. Accounting for vdW interactions and the contraction of the Au(111) surface underneath the $MoS_2$ is further found to be necessary for quantitatively reproducing experimental results. This work demonstrates how the properties of nanostructured $MoS_2$ and similar layered systems can be modified by the choice of supporting metal.

**Introduction**

Materials based on MoS$_2$ and other transition metal dichalcogenides (TMDCs) have a wide range of applications. MoS$_2$ is extensively used as a lubricant[1,2] and as catalyst for the hydrotreatment of fossil fuels[3,4] and the hydrogen evolution reaction (HER).[5] In addition, the electronic and optoelectronic properties of two dimensional (2D) TMDCs[6] have recently attracted considerable interest for their promising applications in the design of nanoelectronic devices.[7,8] The bulk of TMDCs is characterized by layered structures where covalently bonded chalcogen-metal-chalcogen sheets are stacked and bonding through weak van der Waals interactions. Similarly to graphite, which is also used as a lubricant and can be separated as graphene layers, TMDCs can be easily exfoliated due to the weak interlayer interactions, leading to significant changes in the electronic structure. In the single-layer (understood as a S-Mo-S trilayer) limit, MoS$_2$ becomes a direct gap photoluminescent semiconductor,[9,10] in contrast to the indirect gap featured by the bulk crystal. In addition, different phases are known for TMDCs, which essentially differ on the coordination of the metal centers to the surrounding chalcogen atoms and on the stacking pattern of the consecutive layers. In the most stable and common 2H phase for MoS$_2$ (Figure 1a), the Mo atoms are coordinated to the surrounding S atoms in a trigonal prismatic ($D_{3h}$) geometry and layers are stacked in a hexagonal symmetry (ABAB hexagonal stacking). In turn, Mo atoms in the 1T polymorph of MoS$_2$ (Figure 1a) have octahedral coordination ($O_h$) to S and layers are stacked in a tetragonal symmetry (AA stacking). Interestingly, the 1T phase of TMDCs is not only less stable than the semiconducting 2H phase but also metallic.[11] These differences can be rationalized on the basis of band calculations[12] and with simplified representations of the Mo 4d orbitals within crystal field theory. The $D_{3h}$ or $O_h$ coordination of the 2H and 1T phases, respectively, lead to different symmetry-induced splitting of the valence Mo 4d orbitals (Figure 1d). In particular, the $D_{3h}$ symmetry splits the Mo 4d into occupied Mo $4d_{z2}$ and unoccupied Mo $4d_{xz}$, $4d_{yz}$, $4d_{xy}$, and $4d_{x2-y2}$ orbitals, whereas the splitting for $O_h$ symmetry leads to degenerate and semi-occupied Mo $4d_{xy}$ $4d_{xz}$, and $4d_{yz}$ orbitals and unoccupied $d_{z2}$ and $d_{x2-y2}$ orbitals. The lower energy of the fully occupied $4d_{z2}$ orbitals in the 2H phase is responsible for its greater stability and semiconducting character, whereas the semi-occupancy in the 1T phase results in metallic character (Figure 1e). Furthermore, these degenerate semi-occupied states are more accessible to additional electrons than the unoccupied states of the 2H phase. Consequently, the phase transition between 2H and 1T polymorphs can be induced upon different electron donating processes. For example, irradiation with plasmonic hot electrons from deposited Au nanoparticles on MoS$_2$ leads to reversible 2H to 1T phase transitions,[13] and in double-gated field-effect transistors, an insulator to metal transition takes place when reaching critical charge carrier densities.[14] Nevertheless, perhaps the most common method to induce such transitions and stabilize the 1T phase is by doping with alkali metals, which donate their electrons and intercalate as cations between the TMDC layers.[15–17] The resulting materials constitute an attractive electrode material for supercapacitor devices due to the increased hydrophilicity and electrical conductivity of the 1T nanosheets,[18] and have shown increased performance as HER catalysts.[17,19] The activity of MoS$_2$-based catalysts is generally associated with the availability of undercoordinated edge-sites of the layers in MoS$_2$, while the basal plane positions are thought to be more chemically inert.[20] Model systems consisting on supported MoS$_2$ nanoparticles with high edge concentrations have thus typically been used for characterizing the (catalytic) properties of the edges.[21–23] Similarly to the parent 2D (single-layers) and 3D (bulk) structures, MoS$_2$ nanoparticles mostly exhibit the 2H phase and are predominantly terminated in the so-called Mo-edges ($10\bar{1}0$) and S-edges ($\bar{1}0\bar{1}0$), leading to a strong preference for triangular (Figure 1 c and f) and hexagonal nanoparticle shapes.[22,24–26] However, industrial-style MoS$_2$ catalysts preferentially expose the Mo-edges,[23,27] which in contrast to the semiconducting basal plane, have a distinct metallic character that can be distinguished by Scanning Tunneling Microscopy (STM)[28] or X-Ray Photoemission Spectroscopy[29,30] and which is thought to be partly responsible for the increased catalytic activity of such edges.[31] In contrast to the Mo-edge terminations exhibited by large Au(111)-supported MoS$_2$ nanoparticles, small (< 1.5 nm) MoS$_2$ nanoparticles have a preference for S-edge terminations.[22] However, *ab initio* thermodynamics studies involving unsupported MoS$_2$ nanoparticles predict lower edge energies of Mo-edges terminated clusters,[32,33] thereby suggesting that the interaction with the Au(111) support may play an important role in the stability of the different nanoparticle structures. In fact, the interaction with a metallic Au(111) surface has been shown to significantly alter the reactivity of MoS$_2$ stripes due to S bonding to the Au[34] and to affect the band structure of extended MoS$_2$ single layers.[35,36] In another recent study, the chemical and spectroscopic properties of Au-supported nanoparticles were found to depend strongly also on the edge termination exposed, its coverage, and the location (basal plane, edge, or corner) of Mo and S atoms within the MoS$_2$ nanoparticle.[30] Here, we provide a clear understanding of how the interaction with a metallic support affects the stability and properties of different MoS$_2$ nanoparticles. By means of Density Functional Theory (DFT) calculations using explicit size-representative nanoparticle models, we show how support interactions markedly depend on the size, stoichiometry, and edge termination of the of MoS$_2$ nanoparticles.

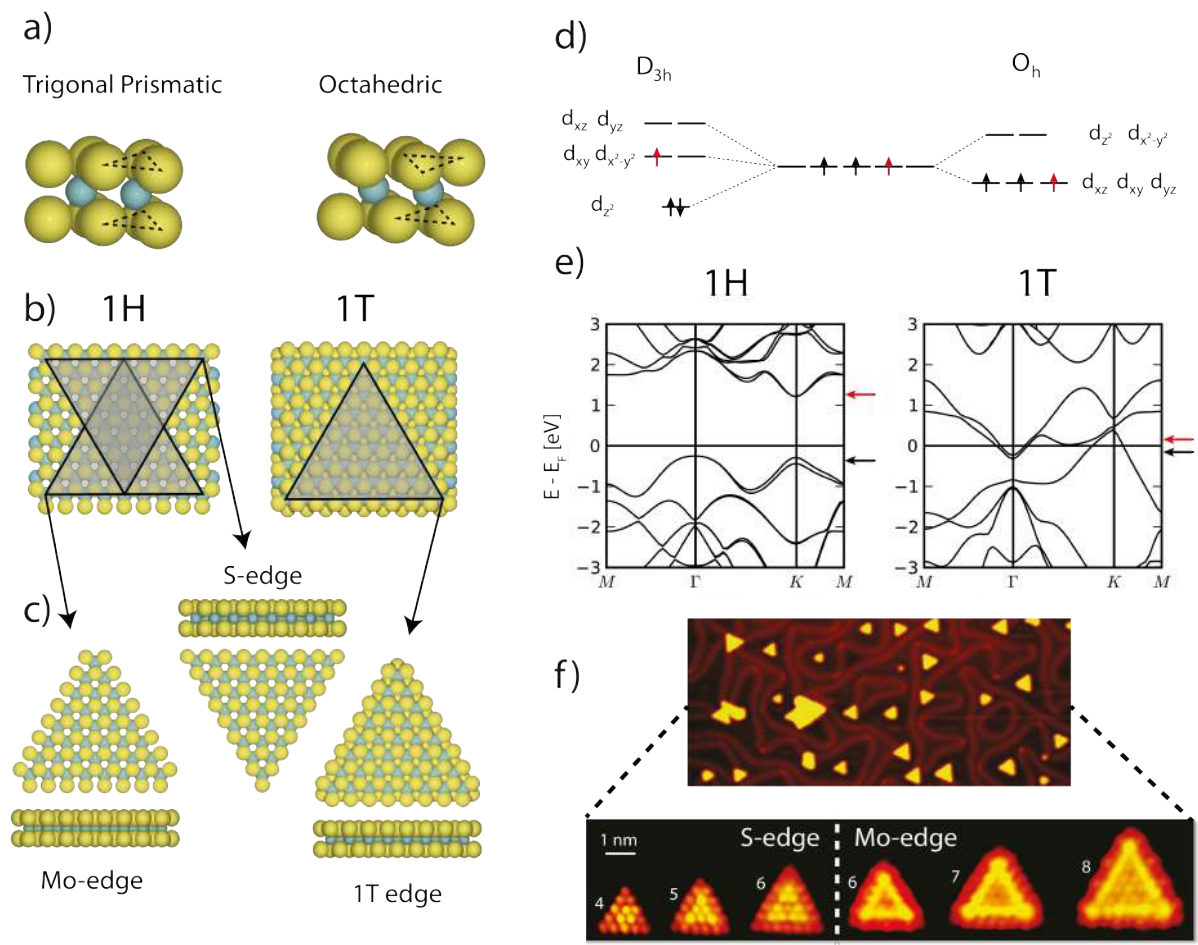

**Figure 1**: (a) Ball models illustrating the trigonal prismatic and octahedric coordination of Mo atoms in the 2H and 1T polymorphs, respectively. (b) Single layers of H- and T- MoS$_2$ with overlaid triangles indicating the shape of the (c) nanoparticle models used in this work. (d) Orbital diagrams for the $D_{3h}$ and $O_h$ symmetries of the metal atom at 2H and 1T phases, respectively, and the corresponding band structure of the resulting single layers. The black and red arrows indicate the energy levels occupied by valence and additional electrons in neutral and electron doped systems, respectively. (f) STM images of triangular MoS$_2$ NPs on Au(111) (adapted from refs [21] and [37]). Individual particles of different size are shown (number of Mo atoms per edge – n – is indicated for each case), illustrating the size threshold for the transition from S-edge to Mo-edge structures.

**Phase diagrams of MoS$_2$ nanoparticles**

We have systematically performed periodic DFT calculations of triangular MoS$_2$ NP models of varying size and edge composition supported on Au(111), see Figure 2. In particular, we have considered NPs with n = 4, 6, 8, and 10, where n corresponds to the number of Mo atoms along each triangle edge. This corresponds to nanostructures with $Mo_{10}S_x$, $Mo_{21}S_x$, $Mo_{36}S_x$, and $Mo_{55}S_x$ stoichiometry, respectively, where the S amount depends on phase and edge S coverage. For each NP size, we have also compared the relative stability of NPs with either trigonal prismatic (2H) or octahedral (1T) coordination of the Mo atoms. For the 2H phase we have considered both S-edge and Mo-edge terminations of the resulting triangular nanoparticles. For each of these NP types (1T, Mo-edge, or S-edge) we have also investigated structures with varying S coverage at the edges. The availability of S at different conditions should be considered when evaluating the relative stability of structures with different stoichiometry. Therefore, we have performed the *ab initio* thermodynamics (AITD) analysis of these models, calculating the Gibbs free energy of formation ($\Delta G$) of the optimized nanoparticle structures upon variations of the chemical potential of S ($\mu_S$) (see methods section for details). This approach allows establishing phase diagrams where the most stable phase at a given chemical potential of S corresponds to the structure with lowest Gibbs free energy of formation $\Delta G$. Therein, S-rich (S-poor) structures are stabilized (destabilized) at high $\mu_S$ (sulfiding conditions), whereas the opposite is true at low $\mu_S$ (reducing conditions).

In Figure 2 we show the calculated phase diagrams for $MoS_2$ NPs of size n = 6. The upper row (a-d) and bottom row (e-h) panels correspond to diagrams of free-standing and Au-supported NPs, respectively. Panels a, b, c, e, f, and g correspond to individual diagrams for each NP type: 2H Mo-edge terminated (blue), 2H S-edge terminated (red), and 1T (green). In these diagrams, each line corresponds to an individual structure, whose slope and intercept depend on its S content and stability, respectively. The most stable phase for each value of $\mu_S$ is therefore the one featuring the lowest $\Delta G$, and the different $\mu_S$ regions are colored accordingly. The last panels of each row (d and h) correspond to the cumulative phase diagrams considering all NP types, where we have plotted the lowest $\Delta G$ values for each NP type along the range of $\mu_S$ and colored according to the most stable NP type.

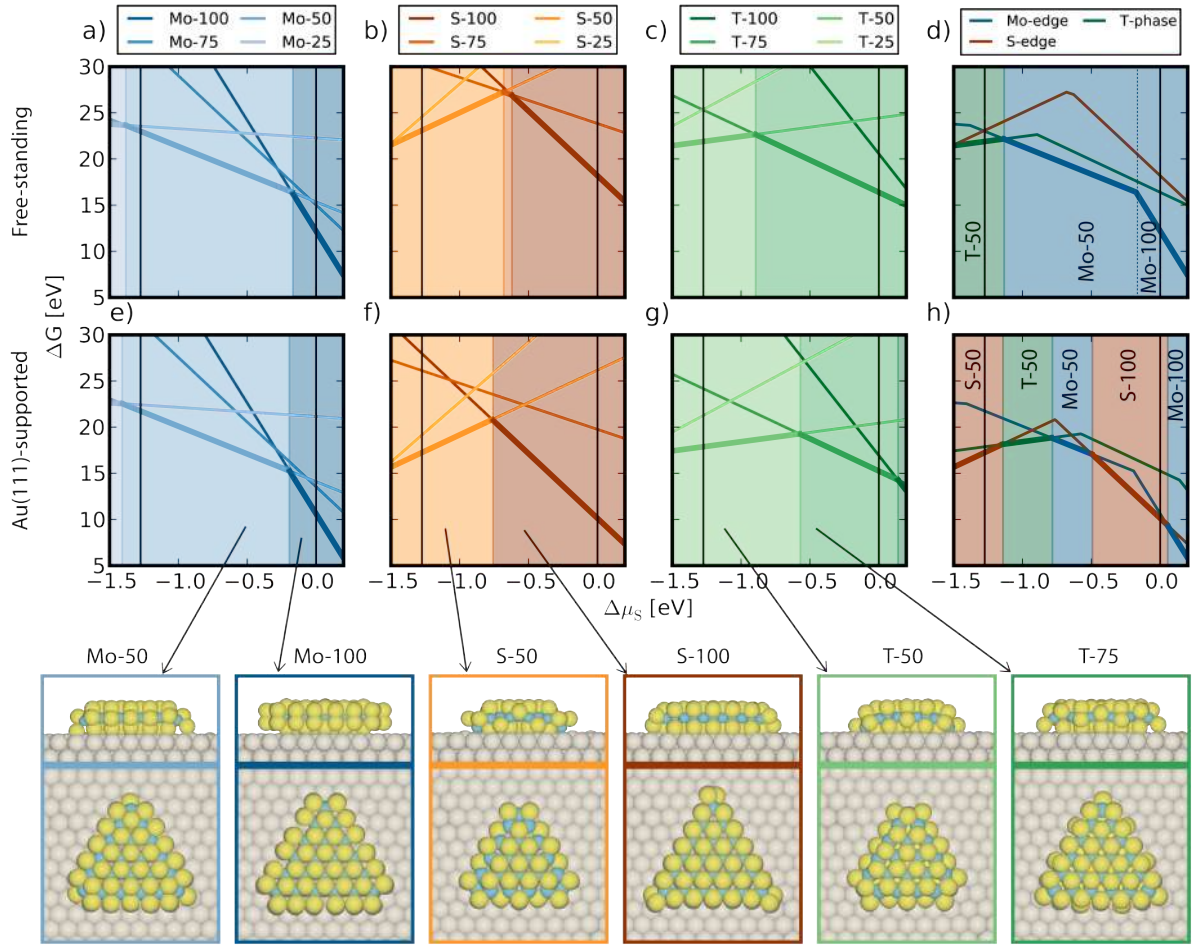

**Figure 2**: Phase diagrams for MoS$_2$ NPs of size n=6 resulting from the *ab initio* thermodynamics analysis of the optimized structures. a-d) Freestanding e-h) Au-supported a,e)Mo-edge b,f) S-edge c,g) T phase d,h) All NP types together. The most stable structures for each NP type at different chemical potentials are shown below the phase diagrams. The labels Mo-, S-, and T-, refer to Mo-edge, S-edge, and T-phase NPs, respectively. The numbers within each label (100, 75, 50, 25) indicate the degree of sulphidation of the edges, in %, with respect the fully sulphided edge (corresponding to the Mo-100, S-100, and T-100 states).

The phase diagram for free-standing Mo-edge particles (Figure 2a) reveals that at strongly sulphiding conditions (high $\mu_S$), the most stable edge configuration is the completely sulphided state (Mo-100, with two terminal S atoms per edge Mo). At lower $\mu_S$ values, this structure is progressively destabilized, and at around $\Delta\mu_S \sim -0.2$ eV the Mo-50 edge configuration becomes more stable. This configuration dominates the entire range of low $\mu_S$ values < −0.2 eV. In turn, Mo-75 and Mo-25 NPs are not favored at any value of $\Delta\mu_S$, or only at potentials below which metallic Mo would form. The situation is similar for the free-standing S-edge particles (Figure 2b), where the fully sulphided S-100 state is the most stable one at high $\Delta\mu_S$ and at ~ −0.6 eV, the S-75 state becomes for a very short range of $\Delta\mu_S$ more stable and dominates under reducing conditions. The preference for Mo-100 and Mo-50 edges and the value of the $\Delta\mu_S$ at which their relative stability is reversed are in agreement with those reported by Schweiger et al.[32] in their study of MoS$_2$ NPs of varying size and composition and coincides well with results obtained using stripe models for describing MoS$_2$ edges.[38,39] The chemical potential at which the transition from free-standing S-100 to S-50 states occurs is also in good agreement with previous work. For the 1T NPs, completely sulphided states – i.e. those preserving all edge S atoms from the bulk cut – are unstable at any range of $\Delta\mu_S$ (Figure 2c). In fact, the oversaturation of S for these structures results in highly distorted structures with reduced crystallinity (not shown).

The compiled phase-diagram including together the three NP types of unsupported MoS$_2$ is shown in Figure 2d. The equilibrium structure at high $\Delta\mu_S$ corresponds to the Mo-100, with a transition to Mo-50 at $\Delta\mu_S \approx -0.2$ eV and another transition to S-50 at $\Delta\mu_S \approx -1.0$ eV. This means that Mo-edge NPs are the most stable for almost the entire range of $\Delta\mu_S$ considered, and only at very low $\Delta\mu_S$, the S-edge NPs become more stable. The clear preference for Mo-edge terminations for free-standing NPs also agrees well with previous studies addressing edge formation energies.[38,39]

Next, the phase diagrams for the supported NPs in Figure 2 (e-h) are analysed, illustrating how the interaction with the Au(111) surface affects the relative stability of the different NP types considered. The phase diagram for Au-supported Mo-edge particles (Figure 2e) is almost identical to that of the freestanding particles and the transition from S-100 to S-50 is found at the same value of $\Delta\mu_S$. This indicates that these NPs interact rather weakly with the metal substrate, which supports that structural and chemical properties of Mo-edges in MoS$_2$ particles prepared on Au supports are not significantly affected by the metal support and therefore constitute appropriate model systems for MoS$_2$-based catalysts.[40] However, the situation for the S-edge is somewhat different and the phase diagram for Au-supported S-edge particles changes with respect to the gas-phase one. In particular, the S-100 particle is the preferred state for a wider range of $\Delta\mu_S$ values and the transition to the S-50 state occurs at lower $\Delta\mu_S$. This is due to the interaction of the S-100 NP with the Au substrate, which is strong enough to even invert the relative stability of S-100 and Mo-100 NPs at high $\Delta\mu_S$ (compare Figure 2d and h). This strong interaction is related to the edge S-atoms of the S-100 state. Each edge S atom in the lower layer (closer to the Au substrate) binds covalently to the

Au atom below (Figure 2, bottom panels), whereas for the Mo-100 NP, upper and lower layer edge S atoms bind to each other preserving the characteristic edge $S_2$ dimers of the freestanding NP. Interestingly, reduced states of the three types of NP interact differently with the Au substrate, leading to a more diverse phase diagram at low chemical potentials. In particular, subsequently lower $\Delta\mu_S$ regions give rise to transitions to Mo-50, a reduced NP of the 1T phase, and finally to S-50 at very low $\Delta\mu_S$. Thus, the reduced T-50 state of the 1T NP type is also stable upon interaction with Au, making this structure the most stable for a particular range of low chemical potentials. The stability of the supported T-50 NP is due to the bonding of edge S atoms to the Au surface and to corner Mo atoms in direct contact with Au, and also to the fact that the free-standing 1T particles are also relatively stable (Figure 2d). These important results show that the interaction with a metallic support will affect the stability of certain terminations or phases of $MoS_2$ NPs and open the possibility of inducing variations of the properties of $MoS_2$ edges via support interactions in combination with reductive treatments. The different interaction with the Au surface of S- and Mo- edges also indicates that in periodic stripe models, which expose both terminations, the properties of the Mo-edge may be affected by the strong interaction of the S-edge side with Au.[34]

**Particle size effects on phase stability**

Intriguing size-effects in Au-supported $MoS_2$ NPs have been identified by means of STM experiments on model catalyst under UHV conditions.[22] In particular, small particles (n ≤ 6) were found to expose S-edge terminations whereas larger particles (n ≥ 6) predominantly expose Mo-edge terminations (Figure 1f). This effect was originally attributed to the oversaturation with S of small Mo-edge NPs, which exceeds the 3:1 S:Mo ratio. For example, the Mo-100 and S-100 NPs for n = 4 correspond to $Mo_{10}S_{36}$ and $Mo_{10}S_{30}$, respectively. In the following lines we demonstrate that the observed preference for S-edge in small Au-supported NPs is instead due to the interaction with the Au support. To elucidate this, we have examined how variations in the size of the $MoS_2$ NPs affect their phase diagrams. We have performed the AITD analysis considering the optimized models of the three kinds of particle type for sizes n = 4 and n = 8, corresponding to smaller and larger particle size, respectively with respect to the diagrams presented in Figure 2. The resulting phase diagrams shown in Figure 3 reveal several differences between these phase diagrams, with stable phases for each NP type appearing at varying $\Delta\mu_S$ values. One of the most notable size effects is related to the stability of the 1T phase. For the smallest particle size considered here (n=4), the 1T is the most stable phase in absence of the Au surface for all allowed values of $\Delta\mu_S$

(Figure 3a), although the Mo-edge NPs are very close in energy. This demonstrates that the Mo-edge termination is more stable than the S-edge termination at high $\Delta\mu_S$ for freestanding particles of even size n = 4, which contradicts the interpretation of size effects based on stoichiometry. As size increases, the free-stranding 1T phase is progressively destabilized with respect to the Mo-edge phase, which becomes the most stable phase already for size n=6 (Figure 3b). The destabilization of the 1T NPs with increase in size is related to larger contribution from basal plane regions, for which the 1T phase is well known to be less stable than the 2H one.[11] These results suggest that 1T $MoS_2$ NPs can be prepared by if particle size is kept small enough and support effects are not too strong; upon interaction with a support such as Au, S-edge NPs become more stable instead (as shown in Figure 3d-f). Preparation methods such as mass-selected deposition[41,42] or gas-phase condensation combined with size-selection and soft landing[43] should therefore allow synthesizing small 1T NPs.

The other relevant size effect appears at the high chemical potential region (around $\Delta\mu_S = 0$) corresponding to sulphiding conditions. As the NP size increases, the value of $\Delta\mu_S$ at which the S-edge termination becomes more stable than the Mo-edge termination is progressively lower. This means that the relative stability between the Mo-100 and S-100 NPs, as well as the chemical potential $\Delta\mu_S$ at which the transition between these two occurs (henceforth referred to as transition potential $\Delta\mu_T$), significantly changes with size. Such variations suggest that the interaction of the S-100 particles with the Au(111) surface weakens as the particle size increases. We further explore the origin of this effect below, but first we address the phase diagrams at lower (more reducing) values of $\Delta\mu_S$. Unlike for n = 6 and 8 NPs, the Mo-50 termination for n = 4 does not become the most stable termination under any conditions within the established $\Delta\mu_S$ limits (indicated by vertical black lines in the phase diagrams). Instead, the preferred termination changes from the S-100 state to the T-50 particle at $\Delta\mu_S \approx -0.45$ eV. Furthermore, in the range from ~−0.45 eV to lower potentials, the S-edge and T-phase structures are almost degenerate, although at $\Delta\mu_S \approx -0.65$ eV the S-50 state becomes more stable. These results indicate that the supported T-phase is relevant for small NPs at reducing conditions, which is not surprising considering that small free-standing 1T structures are already stable and also bind to Au partly through their reduced corners (T-50 structure in Figure 2). The overall stabilizing effect originating from smaller sizes and reduced corner positions is thus less relevant for larger particles, where effects emerging from more abundant basal plane sites dominate.

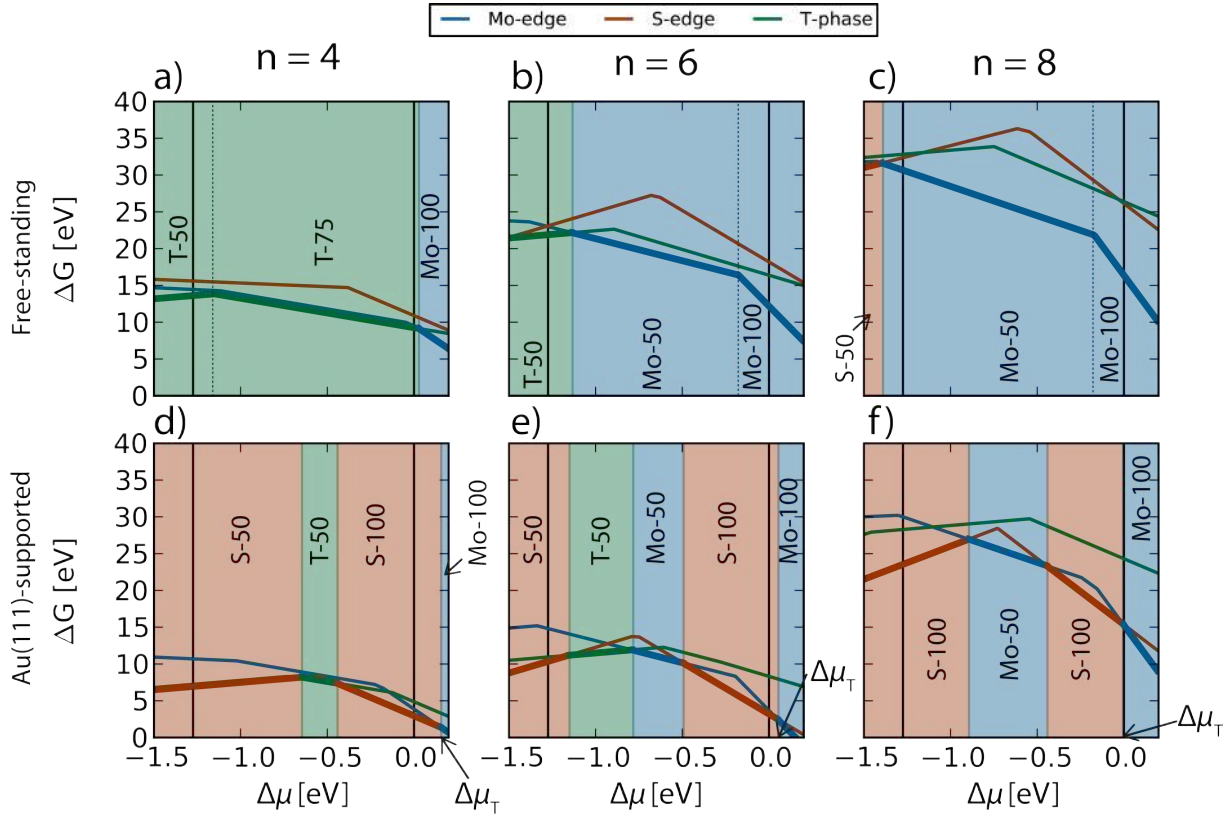

**Figure 3.** Phase diagrams for freestanding (a, b, and c) and Au(111)-supported (d, e, f) MoS$_2$ NPs of sizes n = 4 (a,d), 6 (b,e), and 8 (c,f) resulting from the *ab initio* thermodynamics analysis of the optimized structures. The chemical potential value $\Delta\mu_T$ at which the transition between S-100 and Mo-100 states occurs for the supported nanoparticles is marked. The individual phase diagrams and atomic structures for each NP type of size n = 4 and n = 8 are shown in Figures S1 and S2, respectively.

The variation with size in the value ($\Delta\mu_T$) of chemical potential at which the transition from S-100 to Mo-100 NPs takes place (Figure 3) is indicative of an intricate size effect governing the stability of these nanoparticles. In fact, the STM experiments on the MoS$_2$/Au systems illustrated in Figure 1f revealed that only triangular particles with n ≤ 6 preferentially expose S-edges.[22] The stability of supported NPs is a balance between the formation energy of the freestanding particles and their adhesion energies on Au. In order to elucidate the origin of the variations in relative stability between S-100 and Mo-100 terminations for Au-supported NPs of different size, we have calculated the corresponding formation and adhesion energies for the S-100 and Mo-100 NPs of sizes n = 4, 6, 8, and 10. These quantities correspond to the Gibbs free energies of formation and adhesion at $\Delta\mu_S = 0$, although it is trivial to extend this analysis to the corresponding free energies at other $\Delta\mu_S$ values.

The formation energies of the Mo-100 *and* S-100 freestanding nanoparticles are shown in Figure 4a (blue *and* red bars, respectively), together with the energy gain (black arrows) resulting from their adhesion on Au(111), i.e. their adhesion energy. In this picture, the position of the tip of the black arrowheads therefore indicates the formation energies of the supported particles. The most obvious trend in the calculated formation energies (given per Mo atom in Figure 4a) is that these progressively decrease as particle size increases. This is due to the lower fraction of under-coordinated edge sites present in the larger particles. In fact, the formation energies calculated per edge Mo atom (shown in Figure S3) are almost equal for NPs of different size. With respect to different edge stability, the formation energies for the freestanding S-100 NPs are larger than those of the Mo-100 ones independently of the particle size, in good agreement with previous calculations of the edge formation energy for the two terminations. Most importantly, the adhesion energies of the S-100 particles on Au are much larger (longer black arrows) than for the Mo-edge particles for all the considered particle sizes, leading to a more pronounced stabilization of the S-100 NPs. For the smaller NPs ($n = 4$ and 6) this stabilization is pronounced enough to invert the freestanding relative stability between the two particle types. The interplay between the formation energies of these small freestanding particles and their adhesion energies therefore results in more stable S-100 NPs on Au(111). In contrast, for the larger particles ($n = 8$ and 10) the adhesion energies of the S-100 NPs are not strong enough to overcome the relative energy of the freestanding particles, and the Mo-100 remain as the most stable structures. These results prove that size effects observed in $MoS_2$ NPs are a direct consequence of the interaction with Au(111).

In order to further scrutinize this size effect and clarify why the relative stability is only inverted for the smaller NPs, we have deconvoluted the adhesion energies into three different contributions. In particular, the energy gained upon adhesion of $MoS_2$ MPs on Au(111) can be separated into contributions of bond formation (stabilizing) between S and Au atoms and of the necessary structural deformations of the $MoS_2$ NPs and the Au surface with respect to the unbound situation. These deformations allow stronger adhesion, but also involve a destabilizing energy penalty (see Methods section for further details). Furthermore, one can clarify the role of the different sites of the $MoS_2$ NPs by calculating how the bonding energies scale with respect to the number of such sites. For example, if all atoms at the bottom layer of the $MoS_2$ NPs contribute equally to bond formation, the bonding energy should scale linearly with respect to the number of such S atoms. Alternatively, if edge S atoms dominate the binding of the MoS$_2$ NPs, as we have suggested to be the case for the S-100 NPs, bonding energies should instead scale linearly with number S-edge atoms in contact with the Au.

The bonding energies scaled with respect to the number of lower layer S atoms in each NP (proportional to the area of the NPs) are represented by filled triangles in Figure 4b. These energies are smaller in magnitude (less negative) as size increases, and the variations are much more pronounced for the S-100 particles. However, for the S-100 particles the bonding energy scaled with respect to the number of edge S atoms bound to Au (proportional to the length of the edge), represented by empty triangles in Figure 4b, indeed remains more or less constant for the different particle sizes. This good correlation confirms that the strong bonding of the S-100 structures is mostly due to the binding of the lower-layer edge S atoms to Au, and demonstrates the role of S-edge sites in support interactions with MoS$_2$ nanostructures. Since the strength of the bonds between edge S atoms of the S-100 structures is similar for the different particle sizes, the progressive destabilization of S-100 NPs upon increase in size must originate from larger deformation contributions. The deformation energies of MoS$_2$ NPs are shown in Figure 4c. Deformation energies for the S-100 particles (Figure 4c) are generally larger for the smaller particles. This is consistent with the larger fraction of edge sites, which are distorted more significantly due to their direct participation in the formation of covalent S-Au bonds. However, large S-100 particles are not deformed only at the edges. The bottom panels of Figure 4 show the position and shortest S-Au distance for the S atoms at the bottom layers of the S-100 and Mo-100 NPs. Remarkably, S-100 NPs become increasingly convex as the size increases. This deformation is due to the inherent mismatch between the MoS$_2$ and Au lattices. In particular, the MoS$_2$ particles have a ~9% larger lattice parameter than the Au(111) surface, which means that for completely planar structures edge S atoms are increasingly deviated from positions right on top of an Au atom. These top positions are the preferred sites for Mo-S-Au bond formation, and the supported particles must therefore adopt a convex shape in order to achieve optimal S-Au distances at the edges of all S-100 NPs. Such deformation involves an energy penalty, which for the n = 8 and n = 10 particles results in adhesion energies that are not negative enough to overcome the relative stabilities between freestanding Mo-100 and S-100 structures. We should note again that edge S atoms of the Mo-100 NPs do not participate in the bonding, as clearly indicated by their much larger S-Au distances. The mismatch for these structures results in less pronounced deformations and just slightly less favorable bonding upon increase in size, which in turn leads to a somewhat more separated NPs from the Au surface.

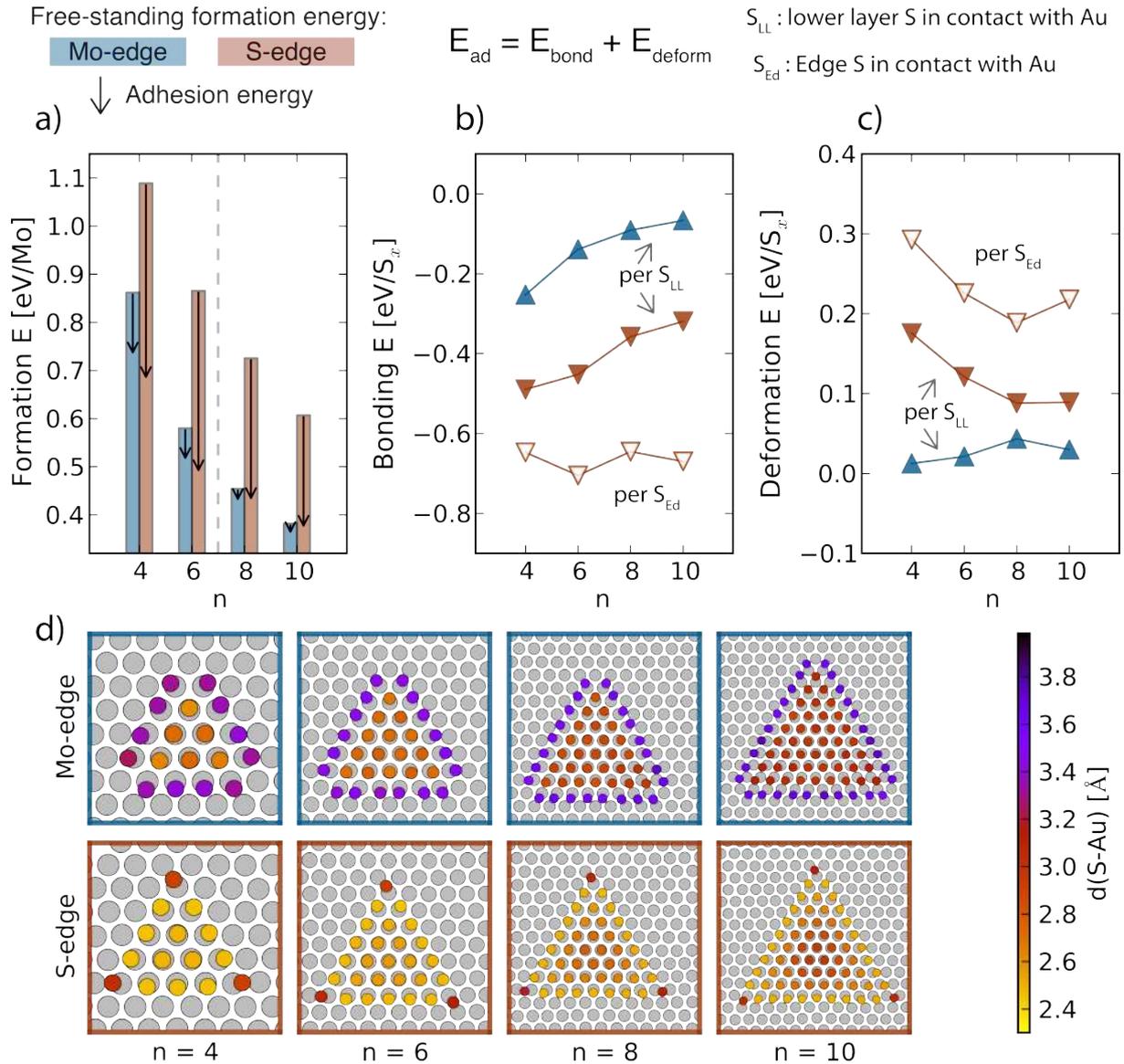

**Figure 4.** Relative stability and structural deformation for Mo-100 and S-100 NPs of sizes n = 4, 6, 8, and 10 on Au(111). a) Formation energies of the freestanding Mo-100 (blue bars) and S-100 (red bars) NPs, with the corresponding adhesion energy on Au(111) indicated by the vertical black arrows. These quantities are normalized with respect to the number of Mo atoms. b) Bonding contribution to the adhesion energies and c) contribution to the adhesion energies from $MoS_2$ NP deformation for Mo-100 (blue) and S-100 (red) NPs. Filled and empty triangles represent energy contributions normalized by the number lower layer S atoms ($S_{LL}$) and edge-S ($S_{Ed}$) in contact with Au, respectively. The bottom panels show the position of S atoms at the bottom layer of the different $MoS_2$ NPs. The S atoms are colored depending on their shortest S-Au distance according to the color-scale at the right.

**Effects of substrate strain and vdW interactions**

We have shown that the relative stability between Mo-100 and S-100 NPs varies with size as a result of an intricate interplay between formation energies of the free-standing particles and the ability of the S-100 NPs to bind through their edge S atoms to the substrate. This explains the experimentally observed preference for either Mo-edge or S-edge terminations in $MoS_2$ NPs of different size. However, we should note that the experimental conditions used during the synthesis of the $MoS_2$ NPs on Au(111) correspond to $\Delta\mu_S = \Delta\mu_{S\text{-syn}} \approx -0.15$ eV (See computational details section). The free energies shown in Figure 3 clearly indicate that at such $\Delta\mu_S$, the S-100 termination is still more stable than the Mo-100 one for the NP sizes considered. This shows that although the weakening of the interaction between S-100 NPs and Au(111) with NP size is well captured by these results, the value of the chemical potential at which the transition from Mo-100 to S-100 phases occurs is not quantitatively reproduced. $MoS_2$ NPs should expose the Mo-100 termination only for particle sizes in which the free energy of formation of the Mo-100 state at $\Delta\mu_S = \Delta\mu_{S\text{-syn}}$ is lower than that of the S-100 NP. This corresponds to a situation in which $\Delta\mu_T < \Delta\mu_{S\text{-syn}}$ and is a condition that should be fulfilled for $MoS_2$ nanoparticles with size n > 6. In order to identify possible sources of this disagreement, we have investigated the sensitivity of our results to the inclusion of vdW interactions in the exchange-correlation functional used and to changes in the lattice parameter used to construct the Au surface model. In particular, we have modeled the Au(111) surface either reproducing the surface compression characteristic of the herringbone reconstruction in Au(111) surfaces or, alternatively, using the optimized lattice parameter of Au (see details in Methods section). We have also used either the PBE *or* optB88-vdW exchange-correlation functionals, whose formulation neglects *or* includes vdW interactions, respectively. Up until now, this work has modeled the $MoS_2$/Au interaction using compressed Au(111) and omitting vdW interactions. In Figure 5, we now show the formation and adhesion energies of the $MoS_2$ nanoparticles resulting from the 4 possible combinations of functional and Au lattice parameter (results in Figure 5b therefore correspond to those of Figure 4a). All approaches yield the same trends, namely, the free-standing Mo-edge particles are more stable, S-edge particles bind more strongly, and both formation energies and adhesion energies decrease (in magnitude) upon increase in NP size. The most significant difference between the different approaches is the offset of the phase-transition potentials $\Delta\mu_T$ with respect to the synthesis potential $\Delta\mu_{S\text{-syn}}$. The furthest agreement from the experimental size threshold corresponds to using the optimized lattice parameter for the Au slab and the PBE functional (Figure 5a). From this situation, either compression of the Au lattice (Figure 5b) or inclusion of vdW interactions (Figure 5c) leads to better agreement with experiment. However, it is only when including both of these effects together (Figure 5d) that the experimental size threshold is reproduced. That is, for sizes n = 4 or 6 (8 or 10) $\Delta\mu_T$ is higher (lower) than $\Delta\mu_{S\text{-syn}}$, which means that S-100 (Mo-100) NPs are more stable. Using the compressed Au slab model therefore makes Au less reactive and weakens its covalent interaction with S-edge MoS$_2$ NPs. In turn, the inclusion of vdW interactions leads to stronger binding for both S-edge and Mo-edge NPs, although the difference is larger for the latter because in absence of vdW interactions these NPs were very weakly bound. Including vdW interactions and considering the compression of the Au atoms in the reconstructed Au(111) surface is therefore necessary for reproducing the experimentally observed particle size threshold.

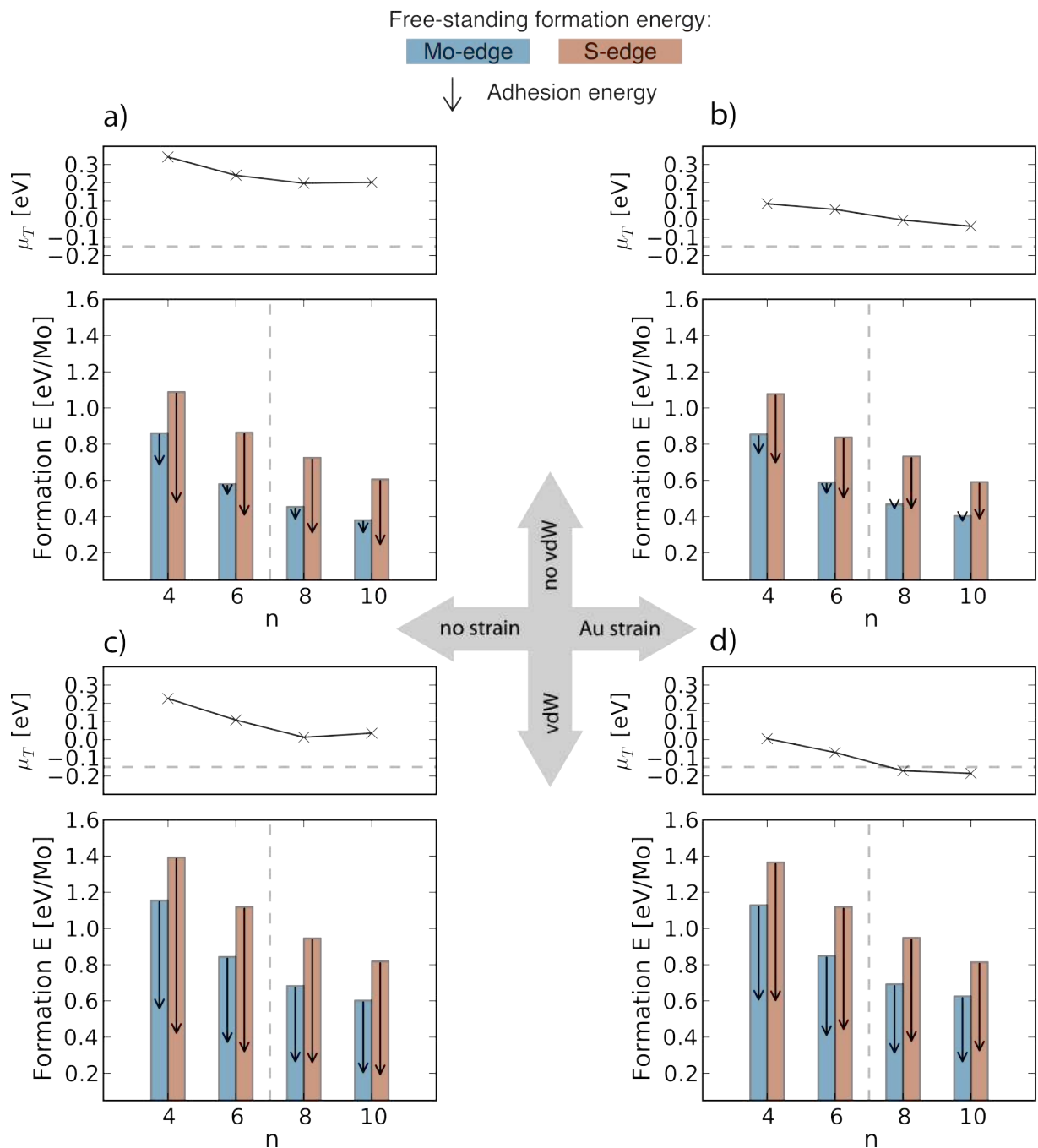

**Figure 5.** Effect of Au(111) surface strain and vdW interactions on the stability of Mo-100 and S-100 NPs of sizes n = 4, 6, 8, and 10 on. Lower panels: Formation energies of the freestanding Mo-100 (blue bars) and S-100 (red bars) NPs, with the corresponding adhesion energy on Au(111) indicated by the vertical black arrows. Upper panels: Evolution of the phase-transition potential $\Delta\mu_T$ with particle size for the different approaches. The horizontal discontinuous line indicates the estimated $\Delta\mu_S$ value for the experimental conditions at which $MoS_2$ nanoparticles are synthesized (−0.15 eV). Panels b correspond to the data presented in Figure 4a.

**Conclusions**

We have shown how particle size and interaction with a metal surface affect the stability of different $MoS_2$ NPs and the resulting phase diagrams. The detailed ab initio thermodynamics analysis presented in this work demonstrates that support interactions stabilize $MoS_2$ terminations and structural phases that are meta-stable for extended monolayers and for free-standing NPs. In particular, the most stable structure for small free-standing $MoS_2$ NPs corresponds to the 1T phase, and these NPs interact strongly at S-poor conditions through reduced corner sites. $MoS_2$ NPs terminated in the S-100 edge are stabilized at S-rich conditions through the formation of strong bonds between edge S atoms and the substrate. These two stabilizing phenomena are more relevant for small NPs; increase in size destabilizes 1T NPs due to the instability of its basal plane. In turn, the mismatch between $MoS_2$ and Au lattices leads to progressively less favorable bonding between S-100 edges and Au upon increases in $MoS_2$ particle size. Size-effects observed experimentally in Au-supported $MoS_2$ NPs can therefore be conclusively attributed to different reactivity of Mo-100 and S-100 edges and to lattice mismatch effects. Because the different phases and edges of $MoS_2$ nanoparticles have different chemical and electronic properties,[30] our results show that $MoS_2$-based nanomaterials can be fine-tuned by support interactions that stabilize $MoS_2$ structures with desirable properties. The dependence of phase stability on the interplay between formation energies, lattice mismatch and edge bonding effect demonstrated here is also useful in general to explain size-effects in other heteroepitaxially grown 2D nanostructures.[44–47]

**Methods**

The DFT calculations of the Au-supported $MoS_2$ nanoparticle models were carried out using the PBE[48] and optB88-vdW[49] exchange-correlation functionals, the projector-augmented wave method of Blöchl,[50] and a real space grid (with spacing of 0.175 Å) for the expansion of the wave functions as implemented in the GPAW code[51] and supported by the Atomic Simulation Environment.[52] The geometries for the different models used were optimized until the forces on each relaxed atom were lower than 0.025 eV/Å and the electronic structure at each geometry optimization step was self-consistently converged with energy, density, and eigenvalue thresholds of 5E−4, 1E−4, and 5E−8 eV, respectively. $MoS_2$ NPs of size n = 4, 6, 8, and 10 were supported on two-layer Au(111) supercells of size 6×5,

8×8, 10×8, and 12×10, respectively. During geometry optimization, all the atoms of the nanoparticles and the upper layer of the Au slab were allowed to relax. In order to obtain converged adsorption energies, for the models on 6×5 *and* 8×8 supercells, total energies were recalculated with single-point calculations using denser 4×4×1 *and* 2×2×1 grids of k-points, respectively. Au(111) slab models were constructed using either the optimized lattice parameter at the level of theory used or a smaller lattice parameter that reproduces the 4

*Phase diagrams and ab initio thermodynamics*: The phase diagrams for the supported nanoparticles were constructed following previous work[32,39,53,54] by calculating the Gibbs free energy of formation (ΔG) as a function of the chemical potential of sulphur. For supported $MoS_2$ nanoparticles:

$$\Delta G(\Delta n, \mu_S) = E_{MoS_2}^{DFT} - n_{Mo} E_{MoS_2}^{ref} - \Delta n \Delta \mu_S - E_{Au}^{DFT}$$

where:

$$\Delta n = 2n_{Mo} - n_S$$

and :

$$\Delta \mu_S = \mu_S - E_{S-bulk}^{DFT}$$

Here, $E_{MoS_2}^{DFT}$ is the DFT-calculated energy of the supported $MoS_2$ nanoparticle, $E_{MoS_2}^{ref}$ is the DFT-calculated energy of the reference bulk $MoS_2$, and $E_{Au}^{DFT}$ is the DFT-calculated energy of the bare Au(111) slab. $n_{Mo}$ and $n_S$ correspond to the number of Mo and S atoms, respectively, of the $MoS_2$. $\Delta n$ therefore represents the variation in stoichiometry with respect to bulk $MoS_2$ of the nanoparticles, which depends on the phase and edge coverage. $\Delta \mu_S$ can take values within the range $\Delta E_F(MoS_2)/2 \leq \Delta \mu_S \leq 0$, for which $MoS_2$ is stable. The lower *and* upper limits define the $\Delta \mu_S$ values at which $MoS_2$ decomposes to metallic Mo *and* S bulk phase is formed, respectively. It is not straightforward to estimate the value of the chemical potential $\Delta \mu_{S-syn}$ corresponding to the conditions at which the $MoS_2$ nanoparticles are synthesized in Au-supported model systems[21–23] because it is difficult to estimate the relative pressure between $H_2S$ and $H_2$. Using T = 673 K and a upper bound value of $p_{H_2S}/p_{H_2} \approx 10^5$, $\Delta \mu_{S-syn}$ is approximated to −0.15 eV as calculated in ref [54].

The adhesion energies have been separated into three different terms: bonding energy, $MoS_2$ deformation energy, and Au deformation energy. The deformation energy corresponds to the destabilizing contribution to the adhesion energy from the structural deformation of the interacting parts upon adhesion. For $MoS_2$ and Au it is calculated as E'($MoS_2$) – E($MoS_2$) and E'(Au) – E(Au), respectively. E'($MoS_2$) and E($MoS_2$) correspond to the energy of $MoS_2$ nanoparticles in their supported structure and in their relaxed free-standing structure, respectively, and the Au terms are defined analogously. The bonding energy ($E_{bond}$) is calculated as: $E_{bond}$ = E(MoS$_2$/Au) – E'(MoS$_2$) – E'(Au), and accounts for the stabilizing interaction between already deformed MoS$_2$ and Au.

## Acknowledgements

We acknowledge support from the Danish Research Council for Independent Research – Technology and Production (HYDECAT, DFF-1335-00016) and Natural Sciences (Sapere Aude Grant no. 0602-02566B), the Innovation Fund Denmark (CAT-C), and the Lundbeck Foundation. AB acknowledges support from the European Research Council under the European Union's Seventh Framework Programme (FP/2007-2013) / Marie Curie Actions / Grant no. 626764 (Nano-DeSign).

## Supporting Information Available

Figures S1, S2 and S3 are included as supporting information. This material is available free of charge via the Internet at http://pubs.acs.org

# Supplementary Information for:

# Size-dependent phase transitions in MoS$_2$ nanoparticles controlled by a metal substrate

Albert Bruix, Jeppe V. Lauritsen, Bjørk Hammer*

Interdisciplinary Nanoscience Center (iNANO) and Department of Physics and Astronomy, *Aarhus University, DK-8000 Aarhus C, Denmark*

*E-mail: hammer@phys.au.dk

This supplementary information document contains 3 figures (S1, S2, and S3).

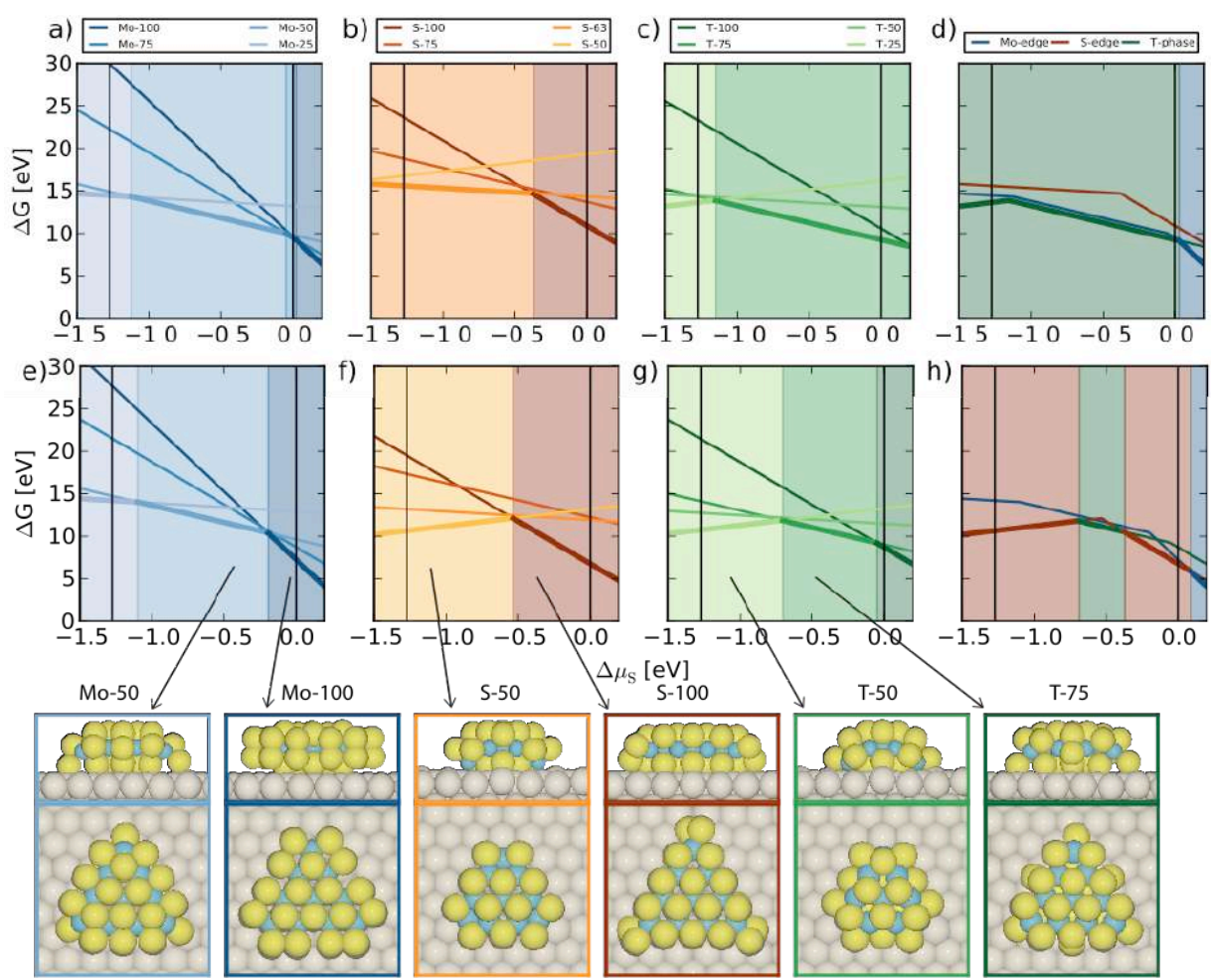

**Figure S1**: Phase diagrams for free-standing (a-d) and Au(111)-supported (e-h) MoS$_2$ NPs of size n=4 resulting from the *ab initio* thermodynamics analysis of the optimized structures.

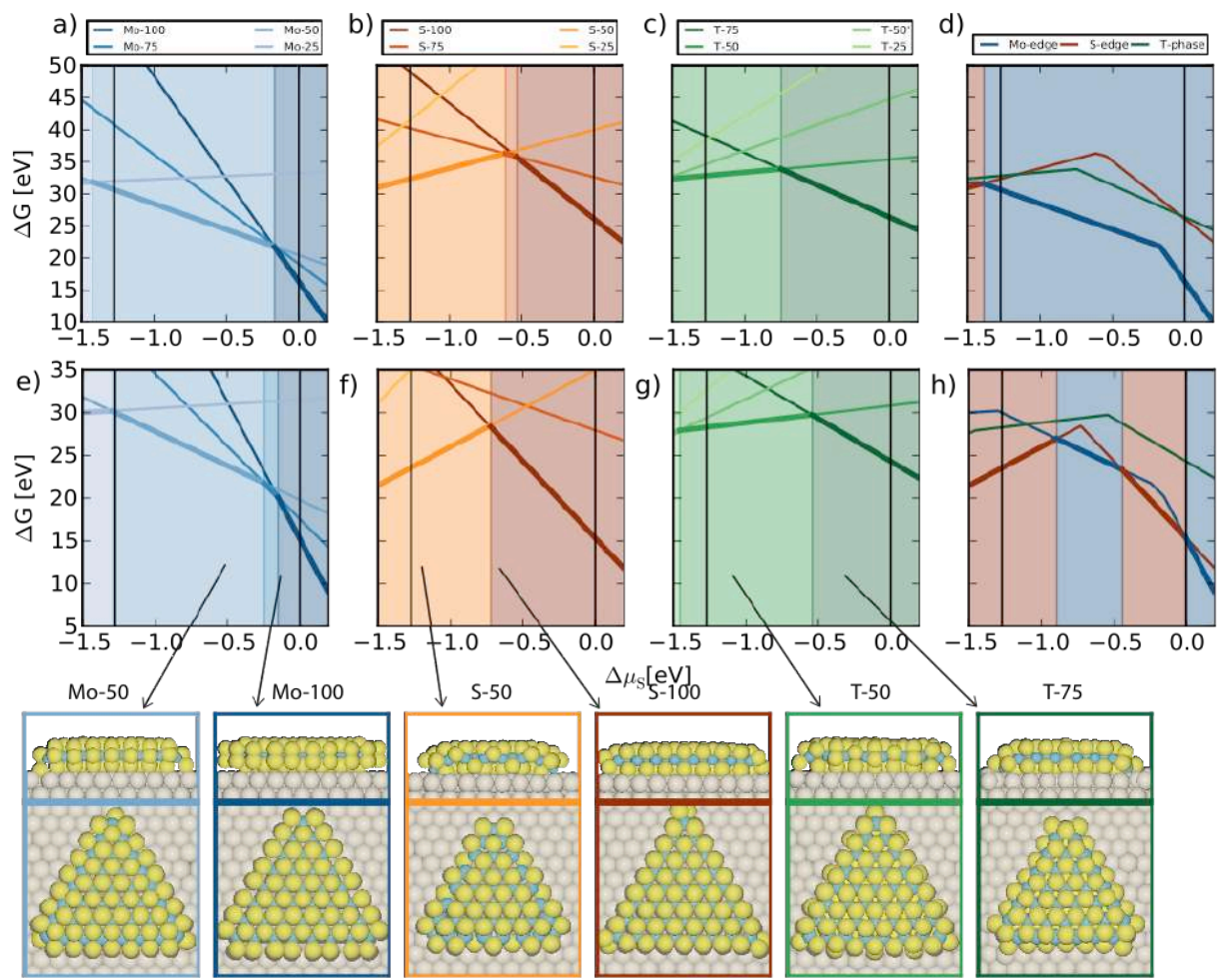

**Figure S2**: Phase diagrams for free-standing (a-d) and Au(111)-supported (e-h) MoS$_2$ NPs of size n=8 resulting from the *ab initio* thermodynamics analysis of the optimized structures.

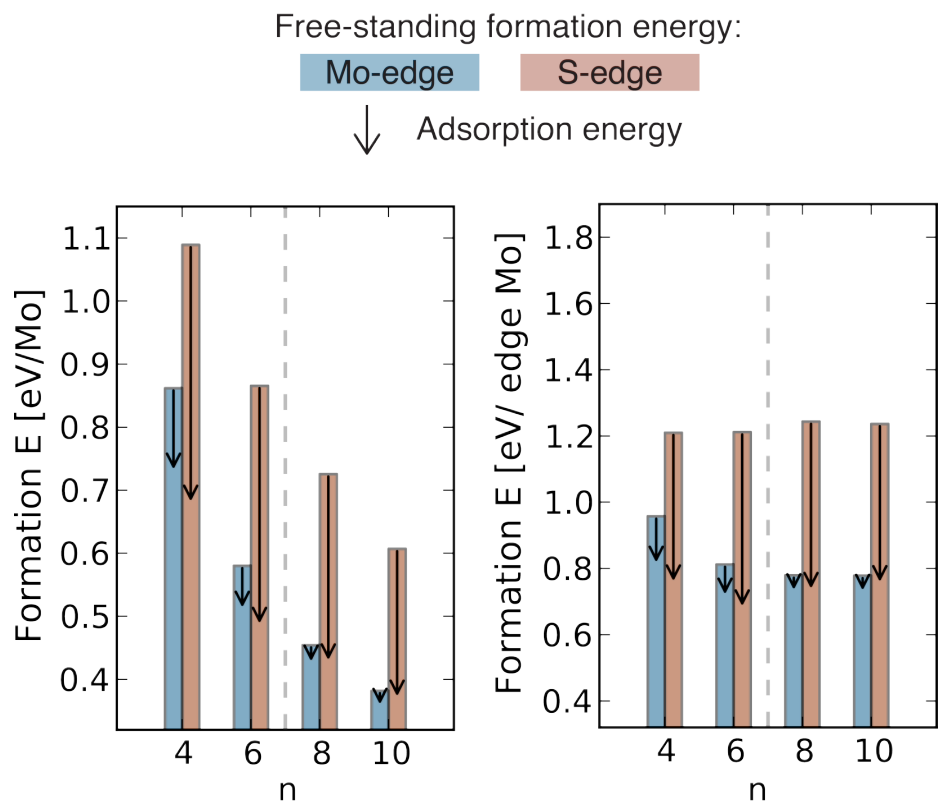

**Figure S3**: Formation energies of the freestanding Mo-100 (blue bars) and S-100 (red bars) MoS$_2$ NPs, with the corresponding adhesion energy on Au(111) indicated by the vertical black arrows. These quantities are normalized with respect to the number of Mo atoms.